# Optimal Feature Selection from VMware ESXi 5.1 Feature Set

Amartya Hatua

Bangalore, India

*Abstract*

*A study of VMware ESXi 5.1 server has been carried out to find the optimal set of parameters which suggest usage of different resources of the server. Feature selection algorithms have been used to extract the optimum set of parameters of the data obtained from VMware ESXi 5.1 server using esxtop command. Multiple virtual machines (VMs) are running in the mentioned server. K-means algorithm is used for clustering the VMs. The goodness of each cluster is determined by Davies Bouldin index and Dunn index respectively. The best cluster is further identified by the determined indices. The features of the best cluster are considered into a set of optimal parameters.*

*Index Terms*

*Clustering, Feature selection, K-means clustering algorithm, VMware ESXi 5.1*

## 1.INTRODUCTION

Feature selection or dimensionality reduction problem for supervised or unsupervised inductive learning basically generates a set of candidate features. These set of features can be defined mainly in three different approaches, a) the number or size of features is specified and they optimize an evaluation measure. b) The subset of features must satisfy some predefined restriction on evaluation measure. c) The subset with the best size and value measured by evaluation measure. Improvement of inductive learning is the prime objective of feature selection. The improvement can be measured in terms of learning speed, ease of the presentation or ability of generalization. To improve the results of inducer, the associated key factors are diminishing the volume of storage, reduction of the noise generated by irrelevant or redundant features and elimination of useless knowledge.

The motivation behind the work comes from selection procedure of VMs during VM migration process or load balancing of VMware ESXi 5.1 server. If a server is overloaded with VMs then some of the VMs should be migrated to other sever. The selection of VMs happens depending on the load of different resources. For every resource there are many parameters or features are present in VMware ESXi 5.1 server. The set of all parameters and their values can be observed using ESXTOP command. Selection of the optimal set of parameters has been carried out in this study. Higher value for particular VM for a set of parameters indicates high usage or load of the particular resource. Depending on the types of load of different VMs; the selection of VMs can be done, which are eligible for migration.

International Journal of Chaos, Control, Modelling and Simulation (IJCCMS) Vol.3, No.3, September 2014

In the following sections Feature selection methods, Clustering methods, Cluster analysis methods, Experiment methodology, Result and Conclusion are elaborately discussed.

## II. FEATURE SELECTION METHODS

A feature selection algorithm (FSA) is a computational solution that is defined by a certain degree of relevance. However the relevance of a feature always depends on the objective of the inductive learning algorithm which uses the reduce set of feature. For induction an irrelevant feature is not useful at the same time all relevant features are not required always [1]. Some of the feature selection methods are: Wrapper method, Filter methods, Rough set methods and Schemata search method. Feature selection using similarity, Principal Component Analysis method, Filter-Wrapper Hybrid method. Again different types filtering method are also available for feature selection, like FOCUS algorithm, RELIEF algorithm, Las Vegas algorithm, Markov Blanket filtering, Decision Tree method and Genetic algorithms [2], [3], [4].

### A. Correlation-based Feature Selection method

Correlation-based Feature Selection (CFS) [5] comes under the category of filter algorithm. In this type of algorithm the subset of features are chosen based on ranks. The rank is determined based on a correlation based heuristic evaluation function. The evaluation function selects features which are highly correlated with the class and at the same time uncorrelated with each other. The evaluation function selects features which are highly correlated with the class and at the same time uncorrelated with each other. Selection of feature happens mainly based on the scale to which it predicts classes in the instance space not predicted by some other features. CFSs feature subset evaluation function is repeated here (with slightly modified notation) for ease of reference:

$$M_s = \frac{k\overline{r_{cf}}}{\sqrt{k + k(k+1)\overline{r_{ff}}}}$$

$M_s$ is representing the heuristic merit of a feature subset $S$ containing $k$ features. $\overline{r_{cf}}$ is denoting the mean feature-class correlation ($f \in s$), and $\overline{r_{ff}}$ is stands for average feature-feature inter correlation. Numerator of the equation is giving an idea regarding the prediction of the class. The denominator is representing redundancy among the features. CFS uses best heuristic search strategy. If continuous five times fully expanded subsets do not give a better result the process stops; in other words that is the stopping condition of the process.

### B. Wrapper method

The wrapper methodology, popularized by Kohavi and John (1997), provides an efficient way to determine the required subset of features, regardless of the machine learning algorithm. Here the machine learning algorithm is considered as a perfect black box. In practice, one needs to define: (i) the procedure to find the best subset among all possible variable subsets; (ii) the procedure to evaluate the prediction performance of a learning machine, which helps to guide the search and reaches the stopping condition; and (iii) the selection of predictor. Wrapper method performs a search in the space of all possible subsets.

Use of machine learning algorithms makes wrapper method remarkably universal and simple. On the other hand when the number of features are very large then this method required an exhaustive search for $O(2^n)$ number of subsets. Where *n* is the number of features. The complexity of the algorithm is $O(2^n)$. So the algorithms is not efficient when *n* is very large.

## C. Filter Algorithm

Filter algorithms determine the subset of optimal features by measuring the intrinsic properties of the data. The algorithm calculates a relevance score for each feature. Features having low relevance score are removed. Afterwards, the subset of relevant features is used as input in different machine learning algorithms.

RELIEF Algorithm: The RELIEF assigns different weight to every feature depending on the relevance of the particular feature on the given context. RELIEF algorithm follows the filter method to determine the minimal set of features. It samples instances from the training data randomly and finds the relevance value based on near-hit and near-miss [6].

## D. Rough Set Method

In Velayutham [7] Rough Set Theory (RST) is used as a tool to discover data dependencies and to find the subset of relevant features.

The rough set is approximation of set by two concepts known as lower and upper approximations. The domain objects are represented by the lower approximation. These domain objects belong to the subset of interest with a certainty. On the other hand the upper approximation describes the objects which are possibly belonging to subset. Approximations are made with respect to a particular subset of features or parameters.

Considering an information system as $I = (U, A \cup d)$. In the information system $U$ is the universe with a non-empty set of finite objects. On the other hand $A$ represents nonempty finite set of condition attributes, and $d$ represents the decision attribute, $\forall a \in A$ the corresponding function is $f_a : U \to V_a$, where $V_a$ represents set of values of $a$. If $P \subseteq A$ then there is an associated equivalence relation:

$$IND(P) = \{(x, y) \in U \times U | \forall a \in P, f_a(x) = f_a(y)\}$$

The partition of $U$ generated by $IND(P)$ is denoted by $U/P$. If $(x, y) \in IND(P)$ holds good, then it is difficult to find a clearly distinguish x and y are by attributes from P. The equivalence classes of the $P$ in discernibility relation are denoted $[x]_P$. Considering $XU$ the P-lower approximation $\underline{P}X$ of set $X$ and P-upper approximation $\overline{P}X$ of set $X$ can be defined as:

$$\underline{P}X = \{X \in U | [X]_P \subseteq X\}$$

$$\overline{P}X = \{X \in U | [X]_P \subseteq X\}$$

Let $P, Q \subseteq A$ be equivalence relations over $U$, then positive region defined as: $POS_P(Q) = \bigcup_{X \in U/Q} \underline{P}X$. The positive region of the partition $U/Q$ with respect to P, $POS_P(Q)$ is the set of all objects of $U$ that can be certainly classified to blocks of the partition . $U/Q$ by means of $P$. Q depends on P in a degree $k (0 \leq k \leq 1)$ denoted $P \Rightarrow_k Q$

$$\text{Where } k = \sigma_P(Q) = \frac{|POS_P(Q)|}{|U|}$$

If $k=1$, Q depends totally on P, if $(0 \leq k \leq 1)$ depends partially on P, if $k=0$ then Q does not depend on P.

The Unsupervised Quick Reduct (USQR) algorithm endeavors to calculate a reduct without fully generating all possible subsets. In Velayutham [8] the algorithm starts off with an empty set. In each iteration one attribute is added in the set. Those attributes are selected depending on the rough set dependency metric. This procedure continues until the value reaches its maximum value for that dataset. The mean dependency of each attribute is calculated and best candidate is chosen:

$$u_P(Q) = \frac{|POS_P(Q)|}{|U|}, \forall a \in A$$

The data set is considered as consistent if the process ends at dependency value 1; otherwise it is inconsistent.

**A. Chi-square feature selection**

Chi-square method calculates the lack of information between a feature and a category. Chi-square distribution is used for comparison with one degree of freedom to judge extremeness [9]. It is defined as:

$$\chi^2(t, c_i) = \frac{N[P(t, c_i)P(\bar{t}, \bar{c_i})P(\bar{t}, c_i)]^2}{P(t)P(\bar{t})P(c_i)P(\bar{c_i})}$$

Here feature selection metrics (four known measures and two proposed variants), which are functions of the following four dependency tuples:

$(t, c_i)$: presence of $t$ and membership in $c_i$.
$(t, \bar{c_i})$: presence of $t$ and non-membership in $\bar{c_i}$.
$(\bar{t}, c_i)$: absence of $\bar{t}$ and membership in $c_i$.
$(\bar{t}, \bar{c_i})$: absence of $t$ and non-membership in $\bar{c_i}$.

# III. CLUSTERING METHOD

K-means clustering [9] is a method of cluster creation, where the main objective is to partition $n$ observations into $K$ groups/clusters in each instance belongs to the cluster with nearest mean. This results in a partitioning of the data space into Voronoi cells. Mathematically Voronoi diagram helps to divide a space into a number of regions. Depending on some predefined conditions set of point is taken in consideration they are known as seeds, sites, or generators. Around each site several points which are closet to that site form a region. The regions are called Voronoi cells. The most popular algorithm uses an iterative technique to reach the stopping point. It is commonly famous as K-means algorithm. In computer science community it is also known as Lloyd's algorithm.

# IV. CLUSTER ANALYSIS METHOD

The cluster analysis methods provide the measure, using which the goodness clusters can be determined. The cluster analysis methods has been used are Davies Bouldin index and
Dunn index (J. C. Dunn 1974).

## A. DaviesBouldin index

The Davies Bouldin index can be calculated by the following formula:

$$DB = \frac{1}{n}\sum_{i=1}^{n} \max_{(x \neq 1)} \left(\frac{\sigma_i + \sigma_j}{d(c_i + c_j)}\right)$$

In the above equation *n* represents the number of clusters. $C_x$ stands for centroid of the cluster *x*. $\sigma_x$ is representing the average distance between each of the point in the cluster and the centroid. $d(c_i + c_j)$ is denoting the distance between two centroid of two clusters.

The clustering algorithm produces a collection of clusters having smallest Davies Bouldin index is considered to be the best algorithm. Clusters with smallest Davies Bouldin index suggests low intra-cluster distances and high intra-cluster similarity.

## B. Dunn index

The main objective of Dunn index is to find dense and well-separated clusters. It is defined as the ratio between the minimal inter- cluster distance to maximal intra-cluster distance. The following formula is used to calculate Dunn index for each cluster partition.

$$D = \min_{0 \leq i \leq n} \left( \min_{1 \leq j \leq n, i \neq j} \left( \frac{d(i,j)}{\max_{1 \leq k \leq n} d'(k)} \right) \right)$$

If *i* and *j* are two clusters then distance between these two cluster is represented by *d(i,j)*. For a cluster *K*, *d(k)* represents intra-cluster distance. The distance between centroids of the clusters can be considered as inter-cluster distance represented by: d(i,j). Similarly intra-cluster distance *d(k)* can be measured as distance between any two elements of cluster *K*. As clusters with high intra-cluster similarity and low inter-cluster similarity are the desirable that is why cluster with high Dunn index are preferable.

## V. EXPERIMENT METHODOLOGY

The steps of the experiment are following:

**Environment setup:** Firstly in a physical server VMware ESXi 5.1 has been installed and twenty six virtual machines (VM) have been installed in that.

**Load selection and run the load:** There are four different types (CPU intensive, Network intensive, Memory intensive and Disk intensive) of load have been run. Each of CPU,
Network and Disk intensive loads has been run in seven different VMs, whereas Network intensive load has been run
in rest of the five VMs.

**Collection of result set:** In third step data has been collected from the server using ESXTOP reply mode. Performance of different factor has been collected like CPU, Memory, Disk, Network and Power.

**Feature selection**: In the result set there are many parameters in this phase the relevant parameters are selected using feature selection algorithms.

**Use K-means algorithm:** By using the clustering algorithms clusters has been created using the subset of features resulting from the feature selection algorithms.

**Comparison:** In this step the clusters has been compared based on some cluster comparison method.

TABLE I. LIST OF FEATURES FOR DIFFERENT RESOURCES

| Resources | Parameters |
|---|---|
| CPU | %USED, %RUN, %SYS, %WAIT, %VMWAIT, %RDY, %IDLE, %OVRLP, %CSTP, %MLMTD, %SWPWT, SWTCH/s, MIG/s |
| Memory | SWCUR, SWTGT, SWR/s, SWW/s, LLSWR/s, LLSWW/s, CPTDR, CPTTGT, ZERO, SHRD,SHRDSVD, COWH, NHN, NMIG, NRMEM, NLMEM, N_L, GST_ND0, OVD_ND0,GST_ ND1, OVD_ND1, OVHDUM, OVHD, OVHDMAX, CMTTGT, CMTTGT, CMTCHRG, CMTPPS, CACHESZ, CACHUSD, ZIP/s, UNZIP/s, MEMSZ, GRANT, SZTGT, TCHD, TCHD W, ACTV, ACTVS, ACTVF, ACTVN, MCTLSZ, MCTLTGT, MCTLMAX |
| Disk | CMDS/s, READS/s, WRITES/s, MBREAD/s, MBWRTN/s, LAT/rd, LAT/wr |
| Network | PKTTX/s, MbTX/s, PKTRX/s, MbRX/s, %DRPTX, %DRPRX, ACTN/s, PKTTXMUL/s, PKTRXMUL/s, PKTTXBRD/s, PKTRXBRD/s |
| Power | %USED, %UTIL, %C0, %C1, %C2, %C3, %T0, %T1, %T2, %T3, %T4, %T5, %T6,%T7 |

# VI.RESULTS

## A.Features selected by using different feature selection algorithms

There are five different feature selection algorithms have been used: CFS method, RELIEF method, Chi Square Method, Wrapper method, Rough set method. The feature selection methods have been used on the data generated from the experiment. The methods select a subset of relevant features depending on some measuring metrics defined by the method.

TABLE I .LIST OF SELECTED FEATURES FOR DIFFERENT RESOURCES

| Resource | CFS | RELIEF | Chi Square | Wrapper | Rough Set |
|---|---|---|---|---|---|
| CPU | %USED, %IDLE, MIG/s | %USED, %RUN, %SYS, %VMWAIT,%RDY, %IDLE, %OVRLP, %SWPWT MIG/s | %USED, %RUN, %SYS, %WAIT, %VMWAIT,%RDY, %IDLE, %OVRLP, %CSTP, %MLMT, | %USED, %RUN, %SYS, %WAIT, %VMWAIT,%RD, %IDLE, %OVRLP, %CSTP, %MLMD, | %USD %SYS, MIG/s |

|  |  |  | %SWPW, SWTCH/s, MIG/s | %SWPT, SWTCH/, MIG/s |  |
|---|---|---|---|---|---|
| Disk | CMDS/s, LAT/r222d | CMDS/s, READS/s, WRITES/, MBREAD/s,LAT/rd, LAT/wr | CDMS/s, READS/s, WRITES/, MBREAD/s,MBWRTN/s,LAT/rd,LAT/wr | CMDS/s, MBREAD/s,LAT/rd | CMDS, LAT/rd, LAT/wr |
| Network | PKTRXs, MbRX/s | PKTTX/s, MbTX/s, PKTRX/s, MbRX/s, ACTN/s, PKTRXMUL/ | PKTTX/s, MbTX/s, PKTRX/s, MbRX/s | PKTTX/s, MbTX/s, PKTRX/s, ACTN/s, PKTRXMUL/ | PKTR/, ACTN/s, PKTRX,MUL/s,PKTRXBRD/s |
| Memory | SWCUR, COWH, OVHD, GRANT, SZTGT | SZTGT, GRANT, MCTLMAX, TCHD W, MEMSZ, TCHD, SHRDSD, ACTVS, MCTLSZ, SHRD, ZERO, SWCUR, MCTLTT, ACTVF, ACTV, CMTTGT, ACTVN, OVHDU, OVHDM, OVHD, COWH, CACHUSD,NLMEM,NHN, NRMEM, N L, CMTPPS, CACHESZ | SWCUR, ZERO, SHRD, SHRDSVD, COWH, NHN, NRMEM, NLMEM, N L, OVHDUM,OVHD, OVHDMAX, CMTTGT, CACHESZ,CACHUSD,MEMSZ,GRANT, SZTGT, TCHD, TCHD W, ACTV, ACTVS, ACTVF, ACTVN, MCTLMAX | GRANT, SZTGT, SWCUR, SHRDSVD,SHRD, ZERO, ACTVS, TCHD W, ACTVF, TCHD, ACTV, CMTTGT, ACTVN, OVHD, MEMSZ, MCTLMAX, MCTLTGT, MCTLSZ, OVHDMAX, OVHDUM, COWH, CACHUSD | SHRDSVD, GRANT |
| Power | %USED | %USED, %C1, %USED | %USED, %C0, %C1 | %USED, %C0, %C1 | %USED,%UTIL,%C1 |

## B.Clustering using K-means algorithm

After getting the selected features K-means clustering algorithm has been applied on each of the set. Where each setconsists of all the selected features of all resources by a particular method of feature selection. Here number of clusteris four. To find the goodness of the clusters, the analysis has been done by the help of Davies Bouldin index and Dunn index.

 Using these two different indices the goodness of the clustering algorithm has been found out. The cluster having maximum Davies Bouldin index and minimum Dunn index is considered to be the best cluster and the feature selection algorithm related to that cluster is considered to be the best algorithm among these algorithms using the data set. From this result it can be concluded that the parameters selected are optimum parameters from which we can have a clear idea about types of load running in VMware ESXi 5.1 server. Here is the table where Davies Bouldin index and Dunn index for different clusters have been given. Where clusters found using the parameters selected by the respective feature selection algorithms.

TABLE I .DAVIES BOULDIN INDEX AND DUNN INDEX FOR CLUSTERS.

| Feature selection technique | Davies Bouldin index | Dunn index |
|---|---|---|
| CFS | 9.9122e+004 | 1.4196e-006 |
| RELIEF | 2.2679e+00 | 1.0874e-009 |
| ChiSquare Method | 2.9018e+007 | 2.442-009 |
| Wrapper | 2.1199e+007 | 1.0862e-009 |
| Rough set | 3.9079e+005 | 4.9040e-006 |

From the result it can be observed that the best result in both of the case is given by the clusters formed by the features selected by the CFS algorithm.

## VII.CONCLUSION

So from the result it can be concluded that the set of parameters selected by the CFS is the best set of parameters of VMware ESXi 5.1. So from that set of parameters we can decide the type of load in VMware ESXi 5.1. At the same time VMs having high values of those parameters suggest the higher usage of those resources by the VM, which can be helpful for load balancing of Virtual Machine.

## REFERENCES

[1] How Useful is Relevance? R.A. Caruana and D. Freitag, Technical report, Fall94 AAAI Symposium on Relevance, New Orleans, 1994, AAAI.
[2] Selection of Relevant Features and Examples in Machine Learning, A.L.Blum and P.Langley, R. Greiner and D. Subramanian, eds., Artificial Intelligence on Relevance, volume 97, pages 245271. Artificial Intelligence, 1997.
[3] An Evaluation of Feature Selection Methods and their Application to Computer Security, J.Doak, An Evaluation of Feature Selection Methods and their Application to Computer Security, 245271, 1992, Technical Report CSE9218, Davis, CA: University of California, Department of Computer Science.
[4] Feature Selection for Knowledge Discovery and Data Mining, H.Liu and H.Motoda, 1998, Kluwer Academic Publishers, London, GB.[5]Mark A.Hall, Feature Selection for Discrete and Numeric Class Machine Learning.
[6] D. W. Yijun Sun, A relief based feature extraction algorithm, Journal of the Royal Statistical Society. Series B (Methodological) Vol. 39, No. 1(1977), pp. 1-38, Tech. Rep., 2011.
[7] A Novel Entropy Based Unsupervised Feature Selection Algorithm Using Rough Set Theory, K. Thangavel and C.Velayutham, Science And Management (ICAESM -2012) March 30, 31, 2012, 2012, IEEE International Conference On Advances In Engineering.
[8] Feature selection for text categorization on imbalanced data, M. A. Hall,Sigkdd Explorations ,Volume 6, Issue 1 - Page 80, Tech. Rep., 2003.K. Elissa, "Title of paper if known," unpublished.
[9] Extensions to the k-Means Algorithm for Clustering Large Data Setswith Categorical Values, ZHEXUE HUANG, P1: SUD Data Mining and Knowledge Discovery KL657-03-Huang 12:59 Data Mining and Knowledge Discovery 2, 283304 (1998) 1998, October 27, 1998, Kluwer Academic Publishers. Manufactured in The Netherlands.